\documentclass[epsfig]{aa}
\usepackage[dvips]{graphics}

\def\ltsima{$\; \buildrel < \over \sim \;$}
\def\simlt{\lower.5ex\hbox{\ltsima}}
\def\gtsima{$\; \buildrel > \over \sim \;$}
\def\simgt{\lower.5ex\hbox{\gtsima}}
 
\begin{document}
   \thesaurus{03 (13.25.2;13.09.1;11.01.2;12.04.2)}
 \title{Optically dim counterparts of hard X-ray selected
AGNs\thanks{Based on observations made with the Italian Telescopio
Nazionale Galileo (TNG) operated on the island of La Palma by the Centro
Galileo Galilei of the CNAA (Consorzio Nazionale per l'Astronomia e
l'Astrofisica) at the Spanish Observatorio del
Roque de los Muchachos of the Instituto de Astrofisica de Canarias.}}

   \author{R.~Maiolino
                \inst{1}
   \and    M.~Salvati
		\inst{1}		
   \and    L.A.~Antonelli
                \inst{2}
   \and    A.~Comastri
                \inst{3}
   \and    F.~Fiore
                \inst{2}
   \and    F.~Ghinassi
                \inst{4}
   \and    R.~Gilli
                \inst{5}
   \and    F.~La~Franca
                \inst{6}
   \and    F.~Mannucci
                \inst{7}
   \and    G.~Risaliti
                \inst{5}
   \and    D.~Thompson
                \inst{8}
   \and    C.~Vignali
                \inst{3,9}
}
   \offprints{R. Maiolino}

  \institute {
 Osservatorio Astrofisico di Arcetri, Largo E. Fermi 5,
I--50125 Firenze, Italy
  \and
 Osservatorio Astronomico di Roma, via Frascati 33,
I-00040, Monteporzio Catone (Roma), Italy
  \and
 Osservatorio Astronomico di Bologna, via Ranzani 1, I-40127, Bologna, Italy
 \and
Telescopio Nazionale Galileo, Aptdo de Correos 565, E-38700 Santa Cruz de La
Palma, Canary Islands, Spain
 \and
 Dipartimento di Astronomia e Scienza dello Spazio, Universit\`a 
 di Firenze, Largo E. Fermi 5, I--50125 Firenze, Italy
  \and
 Dipartimento di Fisica, Universit\`a ``Roma Tre'', via della Vasca
 Navale 84, I-00146 Roma, Italy
  \and
 CAISMI--CNR, Largo E. Fermi 5, I--50125 Firenze, Italy
  \and
 California Institute of Technology, MS 320-47,
  Pasadena, CA, 91125, USA
  \and
 Dipartimento di Astronomia, Universit\`a di Bologna,
 via Ranzani 1, I-40127, Bologna, Italy
}

   \date{Received / Accepted }

\titlerunning{Hard X-ray selected AGNs}
\authorrunning{R. Maiolino et al.}

   \maketitle

   \begin{abstract}
We present near-IR photometry 
and imaging observations of a small sample of sources identified in the
BeppoSAX 5--10 keV survey (HELLAS) which resolves $\sim 20-30$\% of the
X-ray background at these energies.
The near-IR data are combined with optical spectra and
photometry.
Only 40\% of the sources in our sample
 have the blue, power law continuum typical
of color--selected QSOs. The remaining 60\% are dominated by a galactic
component which, on the basis of the continuum colors and shape,
 have ages ranging from
10$^9$ to 10$^{10}$ years. The images show that the blue QSOs are pointlike
at our angular resolution, while all the other sources are extended, 
consistent with their spectral appearance and low redshift. Since down to 
R~=~20 only about two thirds of the HELLAS sources have a counterpart, the 
preliminary HELLAS census comprises in roughly equal parts: i) blue QSOs 
(mostly at high redshifts); ii) optically dim, galaxy--dominated active nuclei
(mostly at modest redshifts); and iii) empty fields (possibly highly
absorbed QSOs at high redshifts).
   \end{abstract}


%

\section{Introduction}

X-ray background (XRB) synthesis models
ascribe most of the
high energy flux to radio quiet,
absorbed Active Galactic Nuclei (AGNs) at intermediate and high redshifts 
(e.g. Comastri et al. 1995, Gilli et al. 1999, Pompilio et al. 2000).
Observationally
the available information on AGNs at these redshifts
 refers mostly to unabsorbed nuclei, 
since the current samples of radio quiet AGNs have been selected mostly with 
color techniques, or in the soft X~rays. Only recently have selection criteria 
less sensitive to absorption been used. Examples are the radio quiet
red QSOs (Kim \& Elvis 1999), analogous to the ones already found in radio 
loud samples (eg. Webster et al. 1995), and the spectroscopic 
identifications in the ELAIS field (Rowan--Robinson et al. 1999). Yet, most 
of our knowledge about absorbed, radio quiet AGNs is limited to low redshifts
and low luminosities, where spectroscopic surveys of bright galaxies have
been performed.

The High Energy LLarge Area Survey [HELLAS, Comastri et al. 2000, Fiore et al.
2000 (paper II)] aims at providing a useful sample of hard X-ray selected
(5--10 keV),
optically identified AGNs while waiting for the Chandra and XMM results. 
The survey instrument is the BeppoSAX MECS.
The sky coverage is 1--50 square degrees at $F_{5-10~keV}
=$~5--30~$\times 10^{-14} \rm erg~cm^{-2} s^{-1}$, respectively, and
is 84 deg$^2$ at fluxes higher than $9\times 10^{-13}
\rm erg~cm^{-2} s^{-1}$. The
cataloged sources amount to 147, and at the fainter limit the source
density is $16.9\pm6.4$~deg$^{-2}$, implying that about 20--30\% of the XRB at
these energies has been resolved.
A program of optical identification is underway, 
which includes optical/near-IR broad band photometry and near-IR
imaging, beside optical spectroscopy of all candidates down to R = 20. 
So far, 63 optical counterparts have been identified, in about two thirds 
of the examined errorboxes. About half of the spectra are typical of QSOs, 
with a blue continuum and broad lines, about half are of intermediate type 
(1.8--1.9), generally with red continua, and a few of them contain only
narrow lines [Fiore et al. 1999 (paper I), La Franca et al. 2000 (paper III)].

In this paper we present and discuss preliminary results of the 
near-IR photometry and imaging observations
of the spectroscopically identified counterparts.
Combined with the optical information presented in papers I and III
these data give a broad-band view of the
properties of the HELLAS sources and allow a preliminary census of
the hard XRB contributors.

\begin{table}[!]
\caption[]{Observing log.}\label{tab_obs}
\begin{tabular}{cccccc}
\hline 
\hline 
\# &\hskip-0.2truecm SAXJ & Date &\hskip-0.2truecm Seeing\hskip-0.4truecm &\multicolumn{2}{c}{Sensit.(T$_{int}$)$^a$} \\
 &  &  &\hskip-0.2truecm($''$)\hskip-0.4truecm &J & Ks \\
\hline 
1  & \hskip-0.2truecm 1118.8+4026A & 19/02/99 &\hskip-0.2truecm 1.0\hskip-0.4truecm & 21.2(4.0) & 19.1(4.0) \\  
2  & \hskip-0.2truecm 2044.6-1028 & 22/07/99 &\hskip-0.2truecm 0.4\hskip-0.4truecm & 21.8(1.6) & 21.0(1.6) \\  
3  & \hskip-0.2truecm 1117.8+4018 & 19/02/99 &\hskip-0.2truecm 0.9\hskip-0.4truecm & 21.3(4.0) & 18.9(2.0) \\  
4  & \hskip-0.2truecm 1118.8+4026B & 19/02/99 &\hskip-0.2truecm 1.0\hskip-0.4truecm & 21.2(4.0) & 19.1(4.0) \\  
5  & \hskip-0.2truecm 0045.7-2515 & 18/12/98 &\hskip-0.2truecm 1.6\hskip-0.4truecm & 20.9(6.3) & 19.0(8.5) \\  
6  & \hskip-0.2truecm 1528.8+1939 & 02/03/99 &\hskip-0.2truecm 1.0\hskip-0.4truecm & --      & 19.3(5.8) \\  
7  & \hskip-0.2truecm 1519.5+6535 & 02/03/99 &\hskip-0.2truecm 0.8\hskip-0.4truecm & --      & 19.4(4.0) \\  
8  & \hskip-0.2truecm 1353.9+1820 & 19/02/99 &\hskip-0.2truecm 1.0\hskip-0.4truecm & 20.8(2.0) & 18.9(3.0) \\  
9  & \hskip-0.2truecm 1218.9+2958 & 19/02/99 &\hskip-0.2truecm 0.9\hskip-0.4truecm & 20.9(2.0) & 18.9(2.0) \\  
10 & \hskip-0.2truecm 1118.2+4028 & 19/02/99 &\hskip-0.2truecm 1.0\hskip-0.4truecm & 20.8(2.0) & 18.8(2.0) \\  
\hline
\end{tabular}
\vskip0.1truecm
$^a$ 3$\sigma$ limiting magnitude (within an aperture twice the seeing) 
 and integration time in minutes.
\end{table}

\begin{table*}[!]
\caption[]{Results.}\label{tab_obs}
\begin{tabular}{lccccccccccccc}
\hline 
\hline 
\# & F$_X^a$ & logN$_H^b$ & L$_X^c$ &
 type$^d$ & z & J$^e$ & Ks$^e$
& \hskip-0.2truecm ext.$^f$ & \hskip-0.2truecm FWHM$^g$ &
 B--R & B--Ks & R--Ks \\
\hline 
1  & 1.2 & $<22.5$ & 45.1 &  B  & 1.13 & 17.98$\pm$0.10 & 16.69$\pm$0.15 & \hskip-0.2truecm N & \hskip-0.2truecm $1''.3$(1.3)   &
 0.42$\pm$0.28 & 1.69$\pm$0.29 & 1.27$\pm$0.29  \\  
2  & 2.0 & $>22.7$ & 46.6 &  B  & 2.76 & 16.25$\pm$0.05 & 14.91$\pm$0.05 & \hskip-0.2truecm N & \hskip-0.2truecm $0''.4$(1.0)    &
 0.43$\pm$0.28 & 2.88$\pm$0.25 & 2.45$\pm$0.25  \\  
3  & 1.3 & $22.7^{+0.5}_{-0.5}$& 45.4 &  B  & 1.27 & 19.38$\pm$0.10 & 17.74$\pm$0.15 & \hskip-0.2truecm N & \hskip-0.2truecm $1''.0$(1.1)    &
 0.55$\pm$0.28 & 2.96$\pm$0.29 & 2.41$\pm$0.29   \\  
4  & --$^h$ & $<22.4$ & --$^h$ &  B  & 0.89 & 19.21$\pm$0.10 & 17.44$\pm$0.15 & \hskip-0.2truecm N & \hskip-0.2truecm $1''.4$(1.4)    &
 0.75$\pm$0.28 & 3.23$\pm$0.29 & 2.48$\pm$0.29    \\  
5  & 3.5 & $22.6^{+0.5}_{-0.5}$ & 43.3 & 1.9 & 0.11 & 16.38$\pm$0.05 & 15.27$\pm$0.07 & \hskip-0.2truecm Y & \hskip-0.2truecm $5''.1$(3.2)    &
 0.80$\pm$0.28 & 2.83$\pm$0.25 & 2.03$\pm$0.25    \\  
6  & 1.4 & $<22.6$ & 43.3 &  L  & 0.18 & --             & 15.94$\pm$0.07 & \hskip-0.2truecm Y & \hskip-0.2truecm $8''.9$(8.9)      &
 1.22$\pm$0.28 & 4.30$\pm$0.25 & 3.08$\pm$0.25    \\  
7  &  11. & $23.2^{+0.1}_{-0.1}$ & 43.2 & 1.9 & 0.04 & --             & 11.38$\pm$0.07 & \hskip-0.2truecm Y & \hskip-0.2truecm $7''.7$(7.7)  &
 1.59$\pm$0.04 & 4.60$\pm$0.08 & 3.01$\pm$0.07     \\  
8  &  8.5 & $22.8^{+0.3}_{-1.2}$& 44.2  &  R  & 0.22 & 15.26$\pm$0.05 & 13.90$\pm$0.07 & \hskip-0.2truecm Y & \hskip-0.2truecm $3''.6$(3.6)     &
 2.18$\pm$0.04 & 5.53$\pm$0.08 & 3.35$\pm$0.07     \\  
9  &  2.4 & $23.1^{+0.4}_{-0.4}$& 43.6 & 1.9 & 0.18 & 16.51$\pm$0.05 & 15.14$\pm$0.07 & \hskip-0.2truecm Y & \hskip-0.2truecm $2''.8$(3.1)    &
  2.35$\pm$0.28 & 5.76$\pm$0.26 & 3.41$\pm$0.26     \\  
10  & 0.9 & $<22.1$ & 43.9 &  R  & 0.39 & 16.48$\pm$0.05 & 14.97$\pm$0.07 & \hskip-0.2truecm Y & \hskip-0.2truecm $2''.9$(2.9)   &
  2.66$\pm$0.28 & 5.83$\pm$0.26 & 3.17$\pm$0.26     \\  
\hline
\end{tabular}
\vskip0.1truecm
$^a$ 5-10 keV flux in units of $\rm 10^{-13}~ erg~ cm^{-2} ~ s^{-1}$;
$^b$ log of intrinsic absorbing column density in units of cm$^{-2}$,
assuming an intrinsic photon index $\Gamma =1.8$ (see paper I) and the
absorber at the redshift of the source;
$^c$ log of the intrinsic (deabsorbed) 5-10 keV luminosity in units
of erg s$^{-1}$, assuming H$_0$=50 and q$_0$=0 (see paper I);
$^d$ optical spectral type: 1.9~-- narrow line
spectrum with broad component of H$\alpha$, L~-- LINER, B~-- broad line QSO
with blue continuum, R~-- broad line QSO with red continuum;
$^e$ Photometry extracted with a circular aperture down to the noise and
corrected for Galactic extinction.
$^f$ Y if resolved in the near-IR images;
$^g$ full-width at half-maximum along the major axis in arcseconds
(quantities in brackets are in seeing units);
$^h$ in the same errorbox of \#1.
\end{table*}

\section{The observations}

Most of the near-IR images were obtained with the ARNICA camera
(Lisi et al. 1996) at the Italian National Telescope Galileo (TNG).
Only one (\#2) out of a total of 10 objects was observed with NIRC (Matthews \&
Soifer 1994) at Keck I. We observed these 10 objects
at K-short ($\sim 2.16 \mu$m) and 8 of them at J ($\sim 1.25 \mu$m).
Tab.~1 gives the observation log 
along with the limiting magnitudes reached in the exposures.
The items are sorted according to the B--R color, which is a measure
of the AGN dominance, as will be discussed in the following.
The first column gives an identification number that, for sake of clarity, 
will be used in place of the full SAX name reported in column 2.
Objects \#1 and \#4 are actually two QSOs that
have been identified within the same errorbox
of one of the HELLAS sources (paper I).

The observations were performed by mosaicing the field every minute,
with offsets of 10-20$''$ around the source,
both to sample the background and to minimize the effects of artifacts
in the array. The data reduction pipeline was similar to that described in
Hunt et al. (1994). Each image was divided by a differential flat field made
out of sky twilight images. Images of each mosaic were aligned by means of
field stars and, then, coadded with a sigma clipping rejection to exclude
hot and dead temporary pixels, not accounted for by the bad pixel mask.

In this paper we also take advantage of the optical spectra used to identify
the sources (papers I and III). In particular we will use the information on 
the continuum shape and the equivalent width of the broad lines.
Optical photometry of these objects was taken from various sources: Palomar 
sky survey, UKST sky survey and some new observations obtained at various 
telescopes in connection with the spectroscopic program.
Each of these optical data set was obtained with slightly 
different filters, however we homogenized the data to the
Johnson system by normalizing the observed spectra to the observed 
photometric points and then resampling in the Johnson bands.

\section{Results and discussion} 

Tab.~2 gives the main results of our near-IR observations. Columns 2 to 4
give the X-ray properties, while columns 5 and 6
give properties inferred from the optical spectra and, in particular,
spectral classification and redshift.
Our subsample of HELLAS sources was not selected with a specific
criterion, it consists of all the early HELLAS identifications known 
and accessible at the time of the observations. This subsample includes
intermediate type (1.9) AGNs, ``classical'' broad line QSOs
with blue continuum, ``red'' broad line QSOs, characterized by a red
underlying continuum, and one LINER, i.e. all the HELLAS identified types.
Our sample also covers the whole range of redshift of the parent sample.
Thus, even if it is not statistically solid because of the small
size, it still contains information on the general properties of the HELLAS
sources. Column 10 gives the full-width at half-maximum of the
near-IR image measured along the major axis. The source is labelled as 
resolved/extended in column 9
if it is at least two times wider than the seeing.

Narrow line AGNs, intermediate type AGNs and red QSOs all show extended
emission indicative of a significant (near-IR) contribution of the host 
galaxy. Furthermore, the B--K colors are not clustered around the value of 
$\sim 2$, typical of color--selected QSOs at high redshift,
but extend to very red values up to B--K$\sim$6, in analogy with the radio
selected AGNs (Webster et al. 1995). Therefore, our red AGNs can be 
considered the radio quiet counterpart of those found in the radio surveys. 
Instead,
all and only blue QSOs are unresolved; as we shall see, this is a consequence
of the dominance of the AGN component and, possibly, of their higher redshift.
These findings are summarized in Fig.~1, where we plot R--K versus B--R and
where point-like and extended sources are marked with circles and squares
respectively.
The solid oblique line is the locus of a single powerlaw, and the 
big cross marks the point where a powerlaw would give B--K~=~2.1.
The dotted line is the reddening curve
for a QSO spectrum\footnote{We used the standard Galactic
extinction curve and a QSO template derived from a
combination of the average spectra given in Elvis et al. (1994)
and Francis et al. (1991).} at $z=0.26$
(the average of the 
three reddest objects, \#8, \#9, and \#10), starting from A$_V$=0.5
for sake of clarity.
The star gives the colors
of an old stellar population\footnote{The galaxy templates were taken from
Bruzual \& Charlot (1993), with solar abundances and a Salpeter IMF.},
again at $z=0.26$.
The dashed line gives
mixed QSO-galaxy contribution models in steps of 10\% relative contribution
(black dots) to the rest-frame V band, down to a minimum galactic contribution
of 50\%.
We see that even the blue QSOs are redder than usual, and the sources with
bent, non--powerlaw spectra are all and only those with extended images.
Also, a QSO template cannot explain all our data, irrespective of the amount
of reddening. Instead, the reddest objects have colors similar to those expected
from an evolved stellar population.
\begin{figure}[!b]
\resizebox{8truecm}{!}{\includegraphics{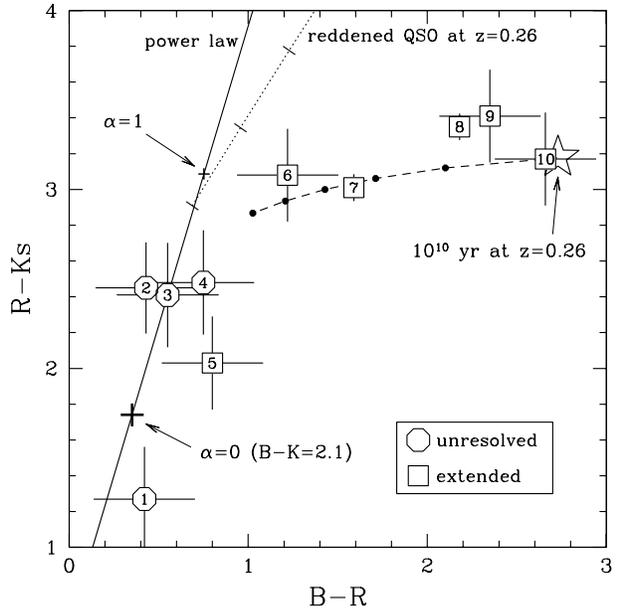}}
\caption{
R--Ks vs B--R color for the sources in our sample (see text).
The numbers refer to the identifications given in Tab.~1.
The dotted line gives the reddening curve of a QSO at z=0.26 and
 starts from A$_V$=0.5;
marks are at A$_V$=0.5, 1.0 and 1.5. The dashed line gives the expected
colors for a model including the contribution of a QSO and an old
stellar population (both at z=0.26); black dots give a galactic
contribution (to the rest frame V-band) of 50\%, 60\%, 70\%, 80\%
and 90\% (the star gives the colors of the stellar template with no AGN
contribution).
}
\label{fig_col}
\end{figure}

As an additional check, we compared the observed optical to near-IR
photometry and optical spectra of each single object
with those expected from (reddened) QSOs and from evolved stellar populations.
An example of this method for a specific object is given in Vignali
et al. (2000).
The combination of the (reddened) AGN and of the galactic component should
give an acceptable fit of the observed photometric points.
We also request that
the model matches the shape of the observed optical spectrum (paper III)
with a maximum tolerance of about 30\% (to allow for uncertainties
introduced by the non-parallactic angle of the slit).
Finally we also tried to fit the equivalent width of the observed broad
hydrogen lines, or to meet their upper limits, within a factor of two
(given the EW spread observed in optical samples of QSOs).
QSO and stellar population templates are the same as for Fig.~1 (notes 1 and 2).
The free parameters of the model are the reddening of the QSO, the age
of the stellar population and the relative contribution of these two
components.
\begin{figure*}[!t]
\centerline{
\resizebox{14.0truecm}{!}{\includegraphics{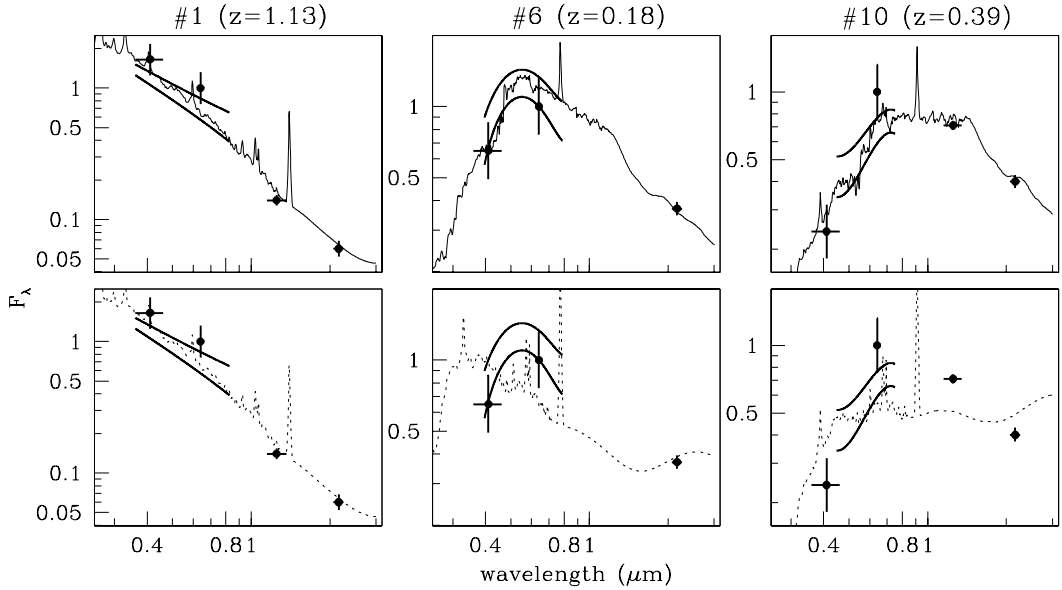}}
}
\caption{Comparison between photometric measurements (black dots with
errorbars) and observed optical
spectral shapes (thick lines)
with the best fit models for three representative objects
(see text). Models in the upper panels (thin solid lines) include both
the contribution from a reddened QSO and from an evolved stellar population;
the latter is dominant in \#6 (10$^9$yr) and in \#10 (10$^{10}$yr).
Lower panels (dotted lines) are the best fit using only a QSO
template with the appropriate reddening.
}
\label{fig_mod2}
\end{figure*}

Fig.~2 shows examples of our spectral fits. They refer to three representative
sources, i.e. the bluest QSO (\#1), a ``transition'' source (\#6), located
around the bend in Fig.~1, and the reddest object of our sample (\#10).
The thin solid line in the upper panels is the
best fit QSO+galaxy model, the two thick lines indicate the shape (and
spread) of the line-free continuum observed in the optical spectra, while
points with errorbars are the photometric fluxes normalized to R.
One sees
clearly the 4000~\AA\ break in the red, spatially resolved sources, with a
discernible progression from the ``transition'' sources to the very red ones.
Along the progression, the B--R color increases from $\sim$1 to $\sim$2.8 and
the preferred age for the model population increases from 10$^9$ to
10$^{10}$ years. The fractional contribution of the reddened
AGN (to the rest frame V-band) decreases from 100\% in the bluest objects to
a few percent in the red ones.
Although the small contribution from
a reddened AGN is required even in the reddest objects, this cannot dominate
their red colors. This is clearly shown by the dotted lines in the lower
panels of Fig.~2,
which give the
best fit to the photometric points using only a (reddened) QSO template: either
the continuum shape or the photometric points, or both, are poorly fitted.
The best model fitting to the photometry and spectral shape is
in perfect agreement with the imaging results, in the sense that 
the contribution of the host galaxy is dominant in those
sources which appear extended. Finally, it is interesting to note that
the best fitting stellar populations are generally old/evolved. However,
one should bear in mind that in certain 
cases different models such as a reddened 10$^9$ year old population or a 
reddened continuous burst provide also an acceptable fit to the data, 
indicating some degree of degeneracy.

Red, absorbed AGNs are about half of the identified sources
in the HELLAS sample, which in turn are about two
thirds of the examined errorboxes.
The fraction of these obscured AGNs is expected
to increase significantly at fainter X-ray fluxes, where the remaining
70--80\% of the hard X-ray background is produced (eg. Gilli 
et al. 1999).
As a consequence, our result suggests that a large fraction of the 
hard XRB contributors have optical/near-IR counterparts which appear as
``normal'' galaxies (possibly with
narrow AGN--like emission lines). 
A population of red AGNs similar to that analyzed in this paper but
at z$>$1 would probably remain undetected at R=20. 
These might be the counterparts
of the HELLAS sources for which no optical identification was found.

Ours are among the first red QSOs selected in hard X-rays, while previous
samples have been selected in the radio (e.g. Webster et al. 1995)
or in the soft X-rays (Kim \& Elvis 1999).
The prevalent interpretation in the latter cases is that the continuum
of the red
QSOs comes from the QSOs themselves, seen through an appropriate
amount of reddening material (eg. Masci et al. 1998, Kim \& Elvis 1999).
In the case of our objects most of the continuum
is instead due to the host galaxy, and absorption is needed
only to make the galaxy's the dominant contribution. Both
Figs.~1 and 2 show that the
reddened QSOs interpretation is untenable, since reddened QSO models fail to
fit colors and spectral shapes.
Also, all of them have extended IR images. The discrepancy with Masci et al.
(1998) and Kim \& Elvis' (1999) results is probably to ascribe to the tendency
of their selection criteria to find QSO that are on average less absorbed
than ours (hence the QSO, although reddened, still dominates over the
galaxy): Kim \& Elvis select their red QSOs among bright soft X-ray sources,
while Masci et al. select flat-spectrum radio sources that, according to the
unified model, should be preferentially seen pole-on. Other studies, which
use selection criteria less sensitive to absorption, as in our case, 
find red QSOs whose continuum is dominated by
their host galaxies, in agreement with what found by us. Among these studies,
Benn et al. (1998) find host galaxy-dominated red QSOs in radio sources which 
have steep radio spectra
(hence preferentially edge-on according to the unified model).
Hasinger et al. (1999) and Lehmann et al. (2000) find several red-AGNs among
faint ROSAT sources whose red colors are ascribed to the
contribution from their hosts; in some of these sources the redshift
moves the rest frame hard-X band into the soft band, while in low-z
objects the depth of the X-ray observation could detect the soft excess
of obscured systems. Finally, Kruper \& Canizares (1989) also found a large
fraction of red-AGNs that are probably dominated by their host galaxies by
selecting sources at X-ray energies (0.5--4.5 keV) higher than ROSAT.
Our finding on red QSOs is in line with the results of Benn et al. (1998),
Lehmann et al. (2000) and Kruper \& Canizares (1989),

\section{Conclusions}

We presented new near-IR (J and Ks band) observations of a sample of 10
objects selected in the hard X-rays (5--10 keV). These sources were
discovered in a large survey (HELLAS) performed by the BeppoSAX satellite,
which resolves $\sim 20-30$\%
of the hard X-ray background. The sample includes 4 blue
broad line QSOs and 6 AGNs with redder continua whose optical
emission line spectra range from broad line objects (red QSOs), to
intermediate type 1.9 AGNs, to LINER.

The B--K color ranges from the standard value of $\sim 2$
(typical of U--B color selected QSOs) up to $\sim$6,
similar to the color of red
QSOs found in radio surveys.

The red AGNs show extended near-IR images. Model fitting
of the photometry and spectral data shows that all and only
the red AGNs are {\it dominated by the emission of the
host galaxy} (with an age of 10$^9$--10$^{10}$ yr).
Red AGNs amount to about a third of the total
HELLAS sources, and their fraction is expected to increase significantly at 
fainter X-ray fluxes, where most of the hard X-ray background
is produced. Therefore, our result suggests that a significant
fraction of the counterparts
of the sources making the hard X-ray background appear as ``normal''
galaxies at optical and near-IR wavelengths.
Chandra and XMM are expected to discover a large number of this
class of objects.

\begin{acknowledgements}
We thank the TNG staff and C. Baffa for technical assistance
during the observations. We are grateful to
P. Giommi, G. Matt, S. Molendi and G.C. Perola, who are involved in the
HELLAS project.
This work was partially supported by the
Italian Space Agency (ASI) through the grant ARS-99-75 and by the
Italian Ministry for University and Research (MURST) through the grant
Cofin-98-02-32.
\end{acknowledgements}

\end{document}